\documentclass[sigconf,screen]{acmart}

\AtBeginDocument{%
  \providecommand\BibTeX{{%
    \normalfont B\kern-0.5em{\scshape i\kern-0.25em b}\kern-0.8em\TeX}}}

 \acmDOI{XXXXXXX.XXXXXXX}

\acmConference[HRI '24]{Workshop on Robo-Identity: Designing for Identity in the Shared World}{March 11, 2024}{Boulder, CO}

\begin{document}


\title{More than just a Tool: People's Perception and Acceptance of Prosocial Delivery Robots as Fellow Road Users}

\author{Vivienne Bihe Chi}
\email{vivienne_chi@brown.edu}
\orcid{1234-5678-9012}
\affiliation{%
  \institution{Brown University}
  \city{Providence}
  \state{RI}
  \country{USA}
}
\author{Elise Ulwelling}
\email{elise_ulwelling@honda-ri.com}
\affiliation{%
  \institution{Honda Research Institute USA, Inc.}
  \streetaddress{70 Rio Robles}
  \city{San Jose}
  \state{CA}
  \country{USA}}
  
\author{Kevin Salubre}
\email{kevin_salubre@honda-ri.com}
\orcid{0000-0002-1410-2556}
\affiliation{%
  \institution{Honda Research Institute USA, Inc.}
  \streetaddress{70 Rio Robles}
  \city{San Jose}
  \state{CA}
  \country{USA}}

\author{Shashank Mehrotra}
\email{shashank\_mehrotra@honda-ri.com}
\orcid{0000-0002-6749-3773}
\affiliation{%
  \institution{Honda Research Institute USA, Inc.}
  \streetaddress{70 Rio Robles}
  \city{San Jose}
  \state{CA}
  \country{USA}}
  
\author{Teruhisa Misu}
\email{tmisu@honda-ri.com}
\orcid{0000-0002-6398-9245}
\affiliation{%
  \institution{Honda Research Institute USA, Inc.}
  \streetaddress{70 Rio Robles}
  \city{San Jose}
  \state{CA}
  \country{USA}}

\author{Kumar Akash}
\email{kakash@honda-ri.com}
\orcid{0000-0003-2807-0943}
\affiliation{%
  \institution{Honda Research Institute USA, Inc.}
  \streetaddress{70 Rio Robles}
  \city{San Jose}
  \state{CA}
  \country{USA}}

\renewcommand{\shortauthors}{Chi, et al.}

\begin{abstract}
  Service robots are increasingly deployed in public spaces, performing functional tasks such as making deliveries.
To better integrate them into our social environment and enhance their adoption, we consider integrating social identities within delivery robots along with their functional identity.
We conducted a virtual reality-based pilot study to explore people’s perceptions and acceptance of delivery robots that perform prosocial behavior. 
Preliminary findings from thematic analysis of semi-structured interviews illustrate people's ambivalence about dual identity. We discussed the emerging themes in light of social identity theory, framing effect, and human-robot intergroup dynamics. 
Building on these insights, we propose that the next generation
of delivery robots should use peer-based framing, an updated value
proposition, and an interactive design that places greater emphasis
on expressing intentionality and emotional responses.
\end{abstract}



\keywords{delivery robot, technology acceptance, prosocial design}


\maketitle
\section{Introduction}
 As service robots become more prevalent in public spaces carrying out tasks such as delivery and cleaning \cite{Wirtz2018,valdez_humans_2023}, people are increasingly having spontaneous interactions with them in dynamic, uncontrolled environments \cite{Astrid2020,Han2023,Bennett_2021}.  Through these interactions, people 
 begin to perceive, relate to, or engage with these robots' social presence beyond their functional role \cite{Clark2022,weinberg_sharing_2023,Nass1995}, allowing these robots to assume dual identities: as functional tools executing designated tasks and as social entities engaging in interpersonal exchanges within the social sphere.  
 
Robot identity plays a crucial role in human-robot interaction, affecting user acceptance, trust, and engagement \cite{Blaurock2022,Miranda2023ExaminingTS}. The prospect of functional robots taking on additional social identities presents both opportunities and challenges. Functional-looking robots, lacking overtly human-like features, could set lower expectations for social intelligence, leading to potential surprise and enhanced acceptability when they effectively perform social tasks \cite{Nazir2023}.
By adopting a newfound identity as social beings, robots traditionally serving single functional roles can expand the range of interactions they can have with other agents---human or robotic---in the shared world, thereby providing added value. 
Moreover, the visibility and exposure of these public-facing service robots may influence not only individual human-robot engagement but also how people act toward one another, with potential impact on social norms and people's wellbeing \cite{winkle_flexibility2021,li2023you,mehrotra2023wellbeing}.
Therefore, for better integration into social and collaborative environments, public-facing service robots need to be portrayed in a more nuanced, sociocultural light beyond their operational functionalities. 
On the other hand, the co-embodiment of functional and social identities raises questions regarding the presentation and transition of robot identities without compromising user trust \cite{Fallatah2019}.
As such, robot designers must construct and present richer and more dynamic robot identities to guide human perception and interaction with them in diverse contexts \cite{Bejarano2022UnderstandingAI}.

In this paper, we consider the case of delivery robots, conceptualizing them not merely as tools but as fellow road users and community members. Despite their functional design, field studies reveal a willingness among laypeople to engage with them socially in the wild \cite{dorrenbacher_towards_2022,sahin_workshop_2021}. However, such interactions have not always been friendly \cite{Oravec2023,Nomura2016,brscic2015,pelikan_encountering_2024}. As road users, delivery robots face a variety of challenges,
where they could benefit from interacting with other human road users \cite{dobrosovestnova_little_2022}.
We propose designing delivery robots that perform prosocial behavior to enhance their social identity.
We conducted a virtual reality (VR)-based study exploring the following two research questions:
 \begin{itemize}
     \item \textbf{RQ1}: How do individuals perceive the prosocial identity of functionally designed delivery robots?
    \item  \textbf{RQ2}: What design changes are necessary to facilitate people's acceptance of delivery robots with dual social-functional identities?
 \end{itemize}
By discussing the implications of our preliminary findings in the following sections, we aim to show how the display of a prosocial identity by delivery robots can foster safety and harmony among all road users as well as boost the social acceptance and adoption \cite{abrams2021,Lim2024} of delivery robots in society.


\begin{figure*}[ht]
    \centering
  \includegraphics[width=.7\textwidth]{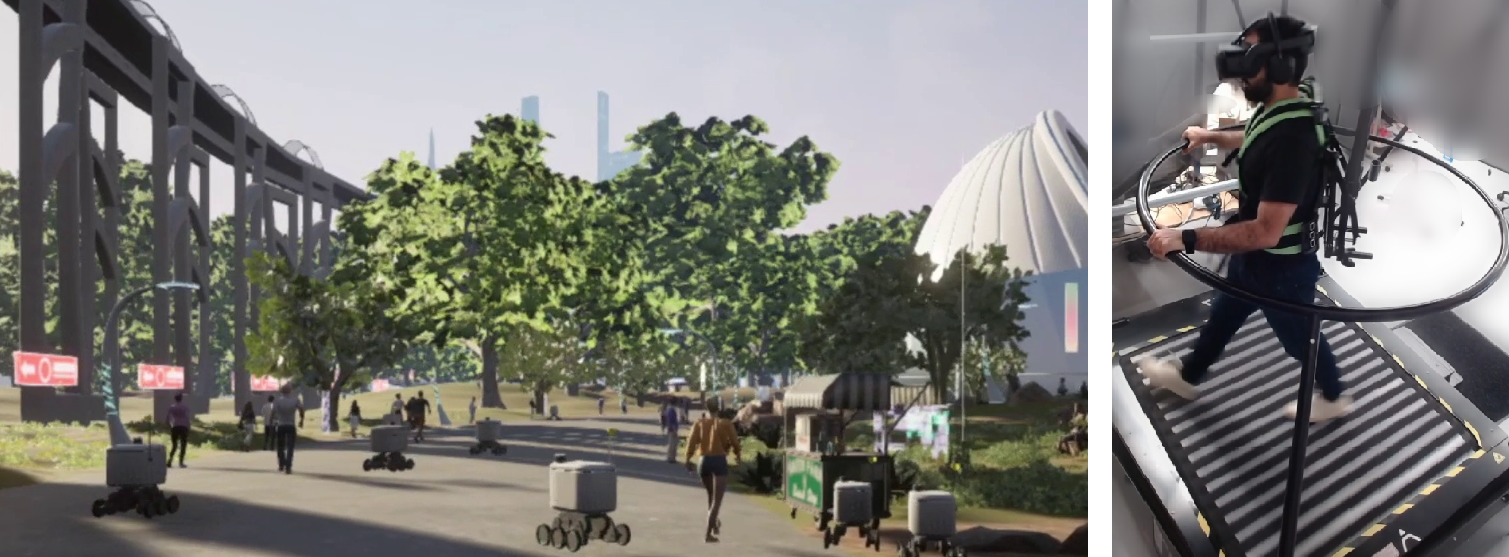}
  \caption{Left: A view of the futuristic environment presented in the virtual reality to the participants. In this city, we show the co-existence of human and AI agents and their prosocial interactions. Right: A participant walking in the virtual environment using an omnidirectional treadmill while wearing a VR headset.}
  \label{fig:simulator}
\end{figure*}
\section{Pilot Study}

\subsection{Materials}
We developed a highly realistic simulation environment utilizing Unreal Engine 5.1.1 \cite{UnrealEngine}, aimed at emulating the futuristic urban landscapes envisioned and advocated in various cities across Europe and North America \cite{cugurullo2021transition}. This environment, as depicted in Figure \ref{fig:simulator}, prominently features autonomously operated delivery robots navigating alongside human pedestrians.Throughout the study, participants encounter about the same number of human pedestrians and delivery robots.
  
Robot identities are primarily conveyed through their behaviors (see Appendix \ref{scripts} for the script used by the researcher to introduce delivery robots to participants). To demonstrate the multifaceted nature of these delivery robots, we curated four distinct on-road scenarios that afford human-robot prosocial interactions:
\begin{enumerate}
\item Warning another road user of oncoming vehicles
\item Notifying another road user of road closures
\item Assisting in the collection of trash misplaced by another road user
\item Helping with a toddler wandered out of parent's sight
\end{enumerate}
Each participant interacted with four instances of each scenario (16 total scenarios). During half of the scenarios, participants witness the prosocial interactions between a pair of fellow road users (robot or human). In the other half of the scenarios, participants have the option to interact with one human or robot in any given scenario.

\subsection{Data Collection}

We recruited 12 participants (9 males and 3 females, 26--35 years)  for an in-person study, which comprised two phases. All participants were employees of Honda Research Institute USA, Inc., San Jose, CA. In the initial phase, participants were immersed in a virtual reality (VR) environment depicting a hypothetical urban setting. Using wireless VIVE Focus 3 VR headset \cite{vive} and Infinadeck omnidirectional treadmill \cite{Infinadeck} (Figure~\ref{fig:simulator}), participants explored the simulation for 40 minutes on average. Throughout this exploration, participants encountered various iterations of the aforementioned scenarios, each unfolding in different global locations and involving different interaction partners, including the delivery robots. Participants were given the freedom to engage with both other pedestrians and the delivery robot during their time in the simulation. 

The subsequent phase of the study involved user interviews. Participants were presented with video snippets recorded during their VR exploration session. Researchers conducted contextual inquiries, focusing on participants' interactions and perceptions within the simulation environment. Open-ended questions were posed regarding participants' beliefs and attitudes toward delivery robots. A comprehensive list of interview questions is provided in the Appendix \ref{appendix}. Research protocols and procedures were approved by Honda's R\&D Bioethics committee (Approval Code: 100HM-020H).

\section{Preliminary Results}
Participants' behavioral responses to the 4 scenarios were recorded. 7 out of 12 participants performed at least one prosocial actions during the experiment session.
Participants' verbal responses were recorded and transcribed for subsequent analysis. Employing a grounded theory-informed thematic analysis \cite{charmaz2014constructing}, we undertook a bottom-up procedure encompassing open coding, axial coding, and the assignment of selective codes, which were then organized into three overarching themes. Below, we present our findings along with relevant participant quotations to elucidate emerging trends. 

\subsection{Robot Capability}
Our initial theme revolves around the capabilities of delivery robots, which were touched upon by participants throughout their responses. Notably, communication capabilities emerged as a focal point, almost all (11/12) expressed felt challenges in communicating with the delivery robots in various situations (``I was just...not sure how to interact with the robot in that situation.'', ``I was testing what interactions were actually possible'', ``even if I tell the robot it[road]'s blocked, it won't understand.'' ), which underscores the importance of bidirectional communication on road.

Concerns regarding the unfamiliarity and perceived limitations of the robots were also raised, with a quarter of the participants (3/12) expressing uncertainty about the robots' capabilities (``Depend on how much I knew about the robot's capabilities. If I knew that the robot could respond to my voice, I might [help].''), while another third (4/12) deemed the robots less capable for their intended undertakings (``I feel more obligated [to act] because I feel like the robot does not have the capacity to.''). Conversely, two-thirds (8/12) of the participants expressed complete trust in the robots' capabilities (``It was very clear that the robot had full perception and should have been able to stop [in this situation].''). It is noteworthy that these judgments were influenced by varying assumptions. Trust, for example, mainly stemmed from the robot's functional capabilities. 

\subsection{Human-Robot Intergroup Relationship}
The second theme delves into the intergroup relationship between humans and robots, reflecting participants' high-level comparisons and perceptions. Half of the participants (6/12) asserted human superiority over robots, emphasizing the intrinsic value of humans over machines (``In that sense, I'm valuing the human's time and energy more than the robot.''). By contrast, a few commented on how humans and robots could be equal (4/12, ``They[humans and robots] had an equal role to play I would say. Because in some situations the robots took care of the like mistakes made by the human or helped users out. The other situations it was the vice versa.'') or acknowledged areas where robots may outperform humans (3/12, ``They[robots] can do tasks at a better success rate than humans.'')

Participants also highlighted the inherent differences between humans and robots, with the majority (9/12) explicitly delineating these disparities (``Because it's programmed... if it[the robot] still fails, that's because it's probably not able to do the task, unlike the human who could have done the task, just chose not to.'').  Specifically, comments centered on differences in mental models (7/12), intentionality (3/12), and predictability (2/12). Moreover, a significant portion (8/12) of participants expressed beliefs that robots operate based on programmed instructions (``The robot is just programmed to do tasks.'').

These observations demonstrated participants' attempts to reconcile their impression of the encountered robots in VR with their preexisting ideas about delivery robots and to form new descriptive and normative opinions on robot identity. Notably, these opinions were occasionally inconsistent even within an individual's response, reflecting the challenges of integrating functional and social identities.

\subsection{Normative Expectations of Robot's Social Identity}
The final theme explores the coexistence of robot identity as a delivery tool and a social agent, reflecting participants' attitudes toward the roles and capabilities of delivery robots. Half of the participants (6/12) exhibited an open attitude toward the social identity of delivery robots (``The robot showed out displayed social responsibility. And I really hope we see a lot of these in the robots coming in the future, that they're not just task-oriented, but they're also socially aware.''), supporting their versatility within the bounds of their capabilities (4/12, ``the more they can do the more they should do.'')

 7 of the 12 participants, representing a somewhat weaker position, contend that robots ought to be specialized or at the very least given priority for functional duties (``Robots are programmed to do their particular tasks...Them being socially aware and being socially intelligent is totally optional.'', ``If it's a delivery robot that's in the process of completing the delivery and in the middle, helping humans to do different kinds of things, I question how effective this delivery robot is regarding completing the task.''). These feelings also stemmed from concerns about robots' core contribution to 
 the community (``the robots should accomplish the tasks. And if they cannot do that, so what are they useful for?'') and one participant's particular mention of moral and regulatory concerns.

Conversely, a small proportion of the participants expressed negative attitudes toward the social identity of delivery robots. They believe that service robots should always follow human commands 
 (4/12), only perform menial, repetitive tasks (4/12) or those difficult for humans to complete (2/12), or they should work autonomously from humans and eliminate the need for human interaction at all (2/12).

\section{Discussion and Future Implications}
The preliminary results underscore the interplay between perceived robot capabilities and the intergroup dynamics between humans and robots, which in turn influenced people's acceptance of the fluid functional and social identities of delivery robots. 

To answer our first research question \textbf{RQ1} about individuals' perception of the prosocial identity of functionally designed delivery robots, the insights derived from the qualitative analysis highlight the active engagement of the three key processes of social identity theory: social categorization, identification, and comparison \cite{Tajfel1979AnIT,Hogg1995,Vanman2019DangerWR}.

Participants appeared to draw on the apparent distinctions between humans and robots, emotional capacity, skill expertise, and transparency of mental models to delineate between `us' (in-group) and `them' (out-group) \cite{Fraune2020}.  This delineation is crucial in understanding the social positioning of robots within human social structures. Further, despite the limited sample size, the current data reveal a tendency among participants to adopt bipolar impressions of robots as either inanimate tools or overly anthropomorphized entities, aligning with previous research \cite{Knight2011}.

The categorization of a robot into a social group triggers a cascade of normative expectations regarding the robot's value and intended purpose \cite{Eyssel2011}. This, in turn, significantly influences an individual's acceptance of the robot's social identity, as evidenced by the thematic analysis of participants' responses. 
    
Participants' ambivalence towards robots' identities, oscillating between tool and social partner, can also be partly attributed to the framing effect \cite{Banks2021,Groom2011,Onnasch2019}. Historical tool-based framing, emphasizing robots' utility in performing mundane and repetitive tasks, persists despite robots demonstrating prosocial behaviors. This underscores the need for a paradigm shift in robot framing---toward a peer-based perspective, enriched with cues like a voice \cite{Callaway2006} and a name \cite{Keay2011}---to foster the ascription of robots' social identity factors \cite{Bejarano2023}.

To address our second research question \textbf{RQ2} about design changes necessary to facilitate people's acceptance of delivery robots with dual social-functional identities, our findings suggest that convincing the public of the value of prosocial delivery robots hinges on demonstrating how they would fit into the community, as the performance of social identity is a collective effort.

Engaging in prosocial acts towards others is an effective identity performance strategy \cite{jackson_design_2021}. In our study, these actions were recognized for their practical value and are particularly appreciated when they set the expectation of reciprocity, reinforcing the significance of reciprocality discussed in the literature \cite{Zonca2021}. 

However, merely observing robot prosocial behavior, as participants did in the simulated environment, falls short of allowing robots to be perceived as peers rather than mere tools. A key insight from our preliminary results is the prevailing skepticism regarding the robots' capacity to fulfill a social-moral role without compromising their performance reliability \cite{Carter2022YoureDM,Walker2020}.
Comparative analysis of participant reactions to prosocial interactions with other human actors and delivery robots revealed that people's evaluations of such encounters are not purely utilitarian. Furthermore, people's motivations for performing prosocial acts extend beyond practical reasons (It feels good to extend kindness and knowing that another human being would appreciate it \cite{Carlson2018}). Participants thus emphasized the difficulty in forming connections with robots, citing communication barriers, the opacity of robots' mental models, and differing expectations of predictability as major hurdles.   

In light of these findings, we advocate for robots to proactively communicate their intentionality \cite{PerezOsorio2019} and display emotional reactions, such as disapproval of norm-violating behavior, frustration when in need, and appreciation after receiving assistance. Implementing these behaviors, particularly in the initial stages of user interaction, could enhance understanding of the robot's objectives and needs, inviting further engagement. Previous studies corroborate the effectiveness of a robot's explicit signaling of its need for assistance \cite{boos2022,MARTIN2020,SANTOS2011} and emotional feedback \cite{Connolly2020} in increasing human willingness to provide help. Such design modifications would foster trust in robots not only regarding their functional capabilities but also in their social-moral dimensions, including benevolence, transparency, and ethicality \cite{MDMT2021}.

Moving forward, we plan to expand the pilot study by continuing to gather data and supplement our qualitative insights with quantitative data obtained from pre- and post-experiment surveys, as well as behavioral measurements recorded during the VR phase of the experiment.

\section{Conclusion}
In our research, we explored the concept of co-embodiment, integrating both functional and social identities within delivery robots, as a strategy to enhance their social acceptance and adoption. Through a VR-based pilot study, we presented and analyzed preliminary results regarding people's perceptions and acceptance of prosocial delivery robots. The results revealed a spectrum of reactions towards the dual identity of delivery robots. By highlighting how robots differ from humans, a majority of participants formed normative expectations for the type of role that these machines should fulfill, posing challenges for simultaneous acceptance of their functional purposes and social attributes. 

Building on these insights, we propose that the next generation of delivery robots should use peer-based framing, an updated value proposition, and an interactive design that places greater emphasis on expressing intentionality and emotional responses.

\section*{Author Biographies}
\textbf{Vivienne Bihe Chi} is a PhD candidate at the Social Cognitive Science Research Lab at Brown University. \\
\textbf{Elise Ulwelling} is a research intern at Honda Research Institute USA, Inc.\\
\textbf{Kevin Salubre} is a human factors researcher at Honda Research Institute USA, Inc.\\
\textbf{Shashank Mehrotra} is a research scientist at  Honda Research Institute USA, Inc.\\
\textbf{Teruhisa Misu} is a chief scientist at Honda Research Institute USA, Inc.\\
\textbf{Kumar Akash} is a senior scientist at Honda Research Institute USA, Inc.\\


\bibliographystyle{ACM-Reference-Format}
\bibliography{robo-identity}

\appendix

\section{Appendix: Semi-Structured Interview Questions}
\label{appendix}
\begin{enumerate}
    \item 	Walk me through your experience today in future city. 
\item	What were you thinking about the pedestrians and the robots? 
\item	\lbrack play VR view video recording\rbrack:  Please describe what happened in this scenario. 
\item	\lbrack for observations\rbrack: What did you think when you saw X helping Y?
\item	\lbrack for decisions\rbrack: How much did you feel obligated to help in this scenario? In this scenario, you made the decision to do X. What was your reason for helping?
\item \lbrack for decisions\rbrack: How much would you feel obligated to help in this scenario if the beneficiary was a [human/robot] instead? 
\item Where there any times that you wanted to help but you didn’t? Why not?
\item Have you experienced receiving help in any of these scenarios in the past?
\item Have you experienced giving help in any of these scenarios in the past?
\item Overall, what do you think about robots helping other humans in this environment?
\item Overall, what do you think about robots helping other robots in this environment?
\item Overall, do you feel that either the delivery robot or the human pedestrians were better suited to helping in these situations? Why? 
\item Any final thoughts about what you saw in the simulation today?

\end{enumerate}

\section{Appendix: Introduction of Delivery Robots}
\label{scripts}
In this experiment, we invite you to future city where you live and work. In future city, autonomous mobilities are highly advanced. They use sensors and smart navigation to reduce potential collisions. You will see delivery robots and smart shuttles that share the streets with pedestrians. You will experience city commutes while wearing AR glasses that add extra information about your surroundings. \\
If you would like to say something to another actor, you can. Like in your daily life, you can interact with the environment with your hands and communicate with other people or robots by speaking.

\end{document}